\documentstyle[11pt]{article}
\newcommand{\rref}[1]{(\ref{#1})}
\newlength{\tablerule}
\def\myrule{\centerline{\rule{\tablerule}{.3pt}}
\centerline{\rule[13pt]{\tablerule}{.3pt}}}
\begin{document}
\vspace*{-1.5cm}
{\obeylines
April 1997\hfill PAR--LPTHE 97-12\\[-5mm]
\hfill ULB--TH 97/08 \\[-5mm]
\flushright hep-th/9704190
}
\vskip 1.5cm
\centerline{\Large \bf On the Opening of Branes}
\vskip1cm
\begin{center}
R.~Argurio${ }^*$\footnote{Aspirant FNRS, e-mail: rargurio@ulb.ac.be}, 
F.~Englert${ }^{*\dagger}$\footnote{e-mail: fenglert@ulb.ac.be}, 
L.~Houart${ }^{* \ddagger}$\footnote{
Charg\'e de Recherches FNRS, e-mail:
lhouart@ulb.ac.be} and P.~Windey${ }^\dagger$\footnote{
e-mail: windey@lpthe.jussieu.fr}\\[.4cm]

${ }^*${\it Service de Physique
Th\'eorique\\ Universit\'e Libre de Bruxelles\\ CP225, Bd. du Triomphe, 1050
Bruxelles, Belgium}\\[.4cm]
${ }^\dagger${\it Laboratoire de Physique Th\'eorique et Hautes Energies
\footnote{ Laboratoire associ\'e No. 280 au CNRS.}\\ 
Universit\'e Pierre et Marie Curie, Paris VI\\ Bte 126,
4 Place Jussieu, 75252 Paris cedex 05, France}\\[.4cm]
${ }^\ddagger${\it Centro de Estudios 
Cient\'{\i}ficos de Santiago\\ Casilla 16443,
Santiago, Chile}
\end{center}
\vskip 1cm

\begin{abstract}
We relate, in 10 and 11 dimensional supergravities, configurations of
intersecting closed branes with vanishing binding energy to configurations
where one of the branes opens and has its boundaries attached to the other.
These boundaries are charged with respect to fields living on the closed
brane. The latter hosts electric and magnetic charges stemming from dual
pairs of open branes terminating on it.  We show that charge conservation,
gauge invariance and supersymmetry entirely determine these charges and
these fields, which can be seen as Goldstone fields of broken
supersymmetry.  Open brane boundary charges can annihilate, restoring the
zero binding energy configuration. This suggests emission of closed branes
by branes, a generalization of closed string emission by D-branes.  We
comment on the relation of the Goldstone fields to matrix models approaches
to M-theory.
\end{abstract}

\newpage
\section{Introduction}

The fact that open strings can end on a D-brane in type II theories of
closed strings \cite{polchinski,pol_lectures} has led to the conjecture
that charged closed $p_a$-branes present in the non perturbative regime can
open and terminate on closed $p_b$-branes. This conjecture is borne out by
the existence, in string theories, of an effective field living on branes:
the latter provides a source term in the $p_b$-brane which can ensure
charge conservation for the opened $p_a$-brane\cite{str}.  An alternative
approach to the charge conservation issue was given by Townsend \cite{tow1}
who showed that the
presence of Chern-Simons type terms in supergravity allows for charge
conservation for well defined pairings of open and closed branes.

On the other hand, general intersection rules between extremal $p$-branes have
recently been derived for type II string theories and
M-theory\cite{papatown,tseytlin_harm,kastor,tse1,arefeva,argu}.  
Intersecting closed branes are
described by classical solutions of supergravity theories with zero binding
energy.  To guarantee the zero binding energy condition for multiple branes
it is sufficient that each intersecting pair satisfies the following
rule\cite{argu}: the intersection of a $p_a$-brane with a $p_b$-brane must
have dimension
\begin{equation}
\bar{q}=\frac{(p_a+1)(p_b+1)}{D-2}-1-\frac{1}{2}\varepsilon_a a_a
\varepsilon_b a_b , \label{intformula}
\end{equation}
where $D$ is the space-time dimension, $a_a$ and $a_b$ are the couplings of
the field strengths to the dilaton in the Einstein frame and
$\varepsilon_a$ and $\varepsilon_b$ are $+1$ or $-1$ if the corresponding
brane is electrically or magnetically charged.  In eleven dimensional
supergravity, the $a$'s are equal to zero while in ten dimension $\varepsilon
a=\frac{1}{2}(3-p)$ for Ramond-Ramond fields and $a=-1$ for the
Neveu-Schwarz three form.

In this paper  we will show that in every case where the zero binding energy
condition is satisfied and where the intersection has dimension $p_a-1$
($p_a\leq p_b$), it is
possible to open the $p_a$-brane along its intersection with the $p_b$-brane.
The list of such intersections is given in tables
\ref{tableiia}-\ref{table11}. The boundary of each open brane configuration
carries a charge living in the closed brane on which it terminates. To each such
configuration corresponds a second one. Their respective boundary charges  are
electric-magnetic dual of  each other in the closed brane and are  coupled to
field strengths in that brane. The fields pertaining to all branes are
all related. They are Goldstone bosons of broken supersymmetry. 

In what follows, we shall first review and apply the
analysis of charge conservation in terms of the  Chern-Simons type terms
\cite{tow1} present in supergravity to the zero binding energy
configurations. We shall then complete this analysis with gauge invariance and
supersymmetry considerations and obtain all world-volume
field strengths and their coupling to boundary charges. 
The significance of these results will be
discussed in the final section.     

\section{Charge conservation from Chern-Simons terms}

In order for a $p_a$-brane of formula (\ref{intformula}) to open along its
intersection with a $p_b$-brane, its boundary must behave, by charge
conservation, as an induced charge living in the $p_b$-brane.  This charge
will act as a source for a field strength on the $p_b$-brane
world-volume. As shown below this field strength is determined
by the Chern-Simons type term present in the supergravity action.  

Firstly we distinguish three different types of configurations ($p_a
\mapsto p_b$) of a $p_a$-brane ending on a $p_b$-brane: (electric $\mapsto$
magnetic), (electric $\mapsto$ electric) and (magnetic $\mapsto$ magnetic),
where by electric brane we mean a brane which couples minimally to a
field of supergravity.  Note that the (magnetic $\mapsto$
electric) never appears since it corresponds to $p_a>p_b$.

Let us first consider the (electric $\mapsto$ magnetic) case in detail.
The equations of motion for the field strength corresponding to the
$p_a$-brane take the generic form
\begin{equation} 
d*F_{p_a+2} = F_{D-p_b-2} \wedge F_{p_b-p_a+1} + Q^e_a
\delta_{D-p_a-1}. \label{emcase}
\end{equation}
Here and in what follows wedge products are defined up to signs and 
numerical factors irrelevant to our discussion.

In the r.h.s. of \rref{emcase} the first term comes from the 
variation of the Chern-Simons term 
\begin{equation}\label{cs}
\int A_{p_a+1}\wedge F_{D-p_b-2}\wedge F_{p_b-p_a+1},
\end{equation}
where $F_{D-p_b-2}$
is the field strength with magnetic coupling to the $p_b$-brane and
$F_{p_b-p_a+1}$ is a $(p_b-p_a+1)$-form field strength present in the
theory. The second term is the $p_a$-brane charge density, where
$\delta_{D-p_a-1}$ is the Dirac delta function in the directions transverse
to the $p_a$-brane.  

The equation \rref{emcase} requires a couple of comments. We want to study
the deformation of intersecting closed brane configurations with zero
binding energy when one of the brane is sliced open along the intersection.
Firstly, notice that we have introduced an explicit source term for the
electric brane since, to study its opening, we want to extend the validity
of the usual closed brane solution on the branes.  This term is required
because the supergravity equations of motion from which the intersecting
brane solutions are derived do not contain any source term and are
therefore valid only outside the sources. Secondly we have to verify that,
in the limit were the open brane closes, the contribution from the
Chern-Simons term vanishes since the closed brane configuration depends
only on $F_{p_a+2}$ and $F_{D-p_b-2}$ and not on the third field
$F_{p_b-p_a+1}$.  The equation for the latter is
\[
d*F_{p_b-p_a+1}=F_{D-p_b-2}\wedge F_{p_a+2}.
\]
It is compatible with a solution where $F_{p_b-p_a+1}=0$ since, when the
two closed branes are orthogonal, the electric $F_{p_a+2}$ and the
magnetic $F_{D-p_b-2}$ have necessarily a common index. 

Taking into account that, in the configuration considered, 
\[
\begin{array}{rcl}
F_{p_b-p_a+1}&=&dA_{p_b-p_a},\qquad\mbox{and} \\
dF_{D-p_b-2}&=&Q^m_b\delta_{D-p_b-1}
\end{array}
\] 
one can rearrange \rref{emcase} in the following way:
\begin{equation}
d(*F_{p_a+2}- F_{D-p_b-2}\wedge A_{p_b-p_a})=Q^e_a \delta_{D-p_a-1} 
		- Q^m_b \delta_{D-p_b-1}\wedge A_{p_b-p_a}.
\label{emcase2}
\end{equation}
The integration of \rref{emcase2} over a $S^{D-p_a-1}$ sphere, which
intersects the $p_a$-brane only at a point and the $p_b$-brane on a
$S^{p_b-p_a}$ sphere which surrounds the intersection, gives
\[
0=Q^e_a- Q^m_b \int_{S^{p_b-p_a}}A_{p_b-p_a}.
\label{chcons} 
\]
This equation can be rewritten as:
\begin{equation}
Q^e_a=Q^m_b . Q_I, 
\qquad \mbox{with} \qquad 
Q_I\equiv\int_{S^{p_b-p_a}}A_{p_b-p_a}.
\label{effcharge} 
\end{equation}
We see that the pull-back $\hat A^{(p_b+1)}_{p_b-p_a}$ of the potential
$A_{p_b-p_a}$ on the closed $p_b$-brane behaves like a $(p_b-p_a)$-form
field strength, magnetically coupled to the boundary.  We will see in the
next section that to preserve gauge invariance one has to add to this field
a $(p_b-p_a)$-form $dW_{p_b-p_a-1}$ defined on the world volume of the
$p_b$-brane.  Although at this stage this is not required, it can be done
without altering the value of the charge $Q_I$. We thus define the field
strength
\begin{equation}\label{calGdef}
{\cal G}_{p_b-p_a}=\hat A^{(p_b+1)}_{p_b-p_a}-dW_{p_b-p_a-1}.
\label{gfield}
\end{equation}
The charge $Q_I$ is then simply a magnetic charge for $\cal G$. 
On the $p_b$-brane ${\cal G}_{p_b-p_a}$
satisfies the  Bianchi identity:
\[
d{\cal G}_{p_b-p_a}=d\hat A^{(p_b+1)}_{p_b-p_a}=\hat F^{(p_b+1)}_{p_b-p_a+1}.
\label{big}
\]
 
For the (electric $\mapsto$ electric) case the reasoning goes along the same
lines starting from the equation of motion:
\[
d*F_{p_a+2} = *F_{p_b+2} \wedge F_{p_b-p_a+1} + Q^e_a \delta_{D-p_a-1}.
\label{eecase}
\]
This time $d*F_{p_b+2} = Q^e_b \delta_{D-p_b-1}$.

For the (magnetic $\mapsto$ magnetic) case we have to consider the Bianchi
identities instead of the equations of motion:
\[
dF_{D-p_a-2}=F_{D-p_b-2} \wedge F_{p_b-p_a+1} + Q^m_a \delta_{D-p_a-1}.  
\label{mmcase}
\]
Here also, we have $dF_{D-p_b-2}=Q^m_b \delta_{D-p_b-1}$.

It is interesting to remark that, as discussed above, the Chern-Simons
terms play no r\^ole in the analysis of intersecting closed branes.  One
sees that requiring the possibility of open brane configurations
necessarily leads to the introduction of these Chern-Simons terms which
from the supergravity point of view are only required by supersymmetry.

\section{Electric-magnetic duality on the brane}

We saw in the preceding section how charge conservation leads to the
existence of a field $\cal G$ whose source is the magnetic charge of the
boundary. However there is an alternative description of this charge. It
arises from the consideration of gauge invariance for the $A_{p_a+1}$
potential and will lead   to the identification of the world-volume
degrees of freedom.

Consider the minimal coupling of the $A_{p_a+1}$ potential to the open
$p_a$-brane.  The boundary of the $p_a$-brane breaks gauge invariance under
$A_{p_a+1} \rightarrow A_{p_a+1}+d\Lambda_{p_a}$ as
\[
\delta \int_{\Sigma_{p_a+1}}A_{p_a+1}= \int_{(\partial\Sigma)_{p_a}}
\Lambda_{p_a}. \label{gaugeinv}
\]
The space-time volume $(\partial\Sigma)_{p_a}$ swept by the boundary of the
$p_a$-brane is the world-volume of a $(p_a-1)$-brane living on the
$p_b$-brane. In order to restore gauge invariance one has to introduce
\cite{witten_bound} a field $V_{p_a}$ living on the $p_b$-brane and
transforming like $V_{p_a}\rightarrow V_{p_a} + \Lambda_{p_a}$.  The
resulting gauge invariant field strength on the $p_b$-brane is
\[
{\cal F}_{p_a+1}=\hat A^{(p_b+1)}_{p_a+1}-dV_{p_a}. \label{ffield}
\]
In string theory, the existence of vector field $V_1$ degrees of freedom
stems from the zero mass excitations of open strings. In the context
of supergravity $V_{p_a}$ emerges from broken supersymmetry.
Indeed, the introduction of a $p_b$-brane breaks half of the
space-time supersymmetries. The broken supersymmetries give rise to eight
massless fermionic degrees of freedom of the $p_b$-brane and their bosonic
partners.  $D-p_b-1$ of them are the Goldstone translation modes of the
$p_b$-brane and are world-volume scalars. The field strength ${\cal
F}_{p_a+1}$ exactly accounts for the remaining Goldstone degrees of freedom.

The charge of the boundary of the $p_a$-brane is measured by
the integral:
\begin{equation}
Q_I=\int_{S^{p_b-p_a}}\star{\cal F}_{p_a+1}, \label{elcharge}
\end{equation}
where $\star$ indicates the Hodge dual on the $p_b$-brane and
the $S^{p_b-p_a}$ encircles the $(p_a-1)$ boundary.
The $Q_I$ in \rref{elcharge} is naturally identified with the one appearing
in \rref{effcharge}.  This identification is in fact required by
supersymmetry, as all massless degrees of freedom have already been
accounted for. Using
\rref{calGdef}, the two expressions for the charge $Q_I$, 
\rref{elcharge} and \rref{effcharge}, imply
\[
{\cal F}_{p_a+1}=\star {\cal G}_{p_b-p_a}. \label{master}
\]
It remains to justify why we introduced a potential  $W_{p_b-p_a-1}$ in 
the definition of $\cal G$ given in \rref{calGdef}. To this end notice that
the Chern-Simons term \rref{cs} could have been written as
\[
\label{cs2}
\int A_{p_b-p_a}\wedge F_{D-p_b-2}\wedge F_{p_a+2}.
\]
In this form the Chern-Simons term is suitable to study the opening of a
$(p_b-p_a-1)$-brane on the $p_b$-brane. This study goes, {\it
mutatis mutandis}, along the same lines as before, with the interchange of
the r\^ole of $\cal F$ and $\cal G$. The introduction of $W$ is now
required to preserve gauge invariance under a transformation of
$A_{p_b-p_a}$, gauge invariance which otherwise would be broken by the
presence of the boundary of the $(p_b-p_a-1)$-brane.
For any $p_b$-brane, open brane configurations always occur in dual pairs.
Their boundary charges are respectively electric 
and magnetic sources for 
$\cal F$, and {\it vice-versa} for $\cal G$.

To summarize, the  equations for the gauge field living on the
$p_b$-brane are in general 
\[
\begin{array}{rcl}
d\star {\cal F}_{p_a+1}&=&\hat F^{(p_b+1)}_{p_b-p_a+1},\\
d {\cal F}_{p_a+1}&=&\hat F^{(p_b+1)}_{p_a+2}.
\end{array} \label{feqs}
\]

Below we give a complete list of the  intersections which give rise to open
brane configurations for IIA, IIB and eleven dimensional supergravity.

The bosonic field content of IIA supergravity consists of the fields
\[
\label{nsfields}
(\phi,g_{\mu\nu}, B_2)\ ,
\]
coming from the NS-NS sector of string theory, together with the
gauge potentials from the RR sector of IIA
\[
(A_1, A_3).
\label{2arfields}
\]
The 4-form field strength which appears in the equations of motion and
Bianchi identities is given by $F^{\prime}_4=dA_3+A_1\wedge H_3$.

For type IIB supergravity the bosonic field content is, besides the common
NS-NS sector, the gauge potentials of the RR fields
\[
(A_0,A_2,A_4),
\label{2bfields}
\]
where $A_4$ is a potential related to the self-dual 5-form field strength $F_5$.
We define $F^{\prime}_3=dA_2-A_0 H_3$.

In 11 dimensional supergravity there are only two bosonic fields
\[
\label{mfields}
(g_{\mu\nu}, A_3).
\]

In the tables we list in each row the intersection, the relevant equation
of motion or Bianchi identity, and the equations for the world-volume field
strength.  The configurations are arranged by dual electric-magnetic pairs.
\begin{table}[h]
\myrule\vskip -.5cm
\[
\begin{array}{ccc}
\bf{1^e_F\mapsto 2^e_D}
	&\qquad d*H_3=*F^{\prime}_4\wedge F_2
	&\qquad d\star {\cal F}_2 =\hat F^{(3)}_2\\
\bf{0^e_D\mapsto 2^e_D}
	&\qquad d*F_2=*F^{\prime}_4\wedge H_3
	&\qquad d{\cal F}_2 =\hat H^{(3)}_3\\ 
& & \\ 
\bf{1^e_F\mapsto 4^m_D}
	&\qquad  d*H_3= F^{\prime}_4\wedge F^{\prime}_4
	&\qquad d\star {\cal F}_2 =\hat F^{\prime (5)}_4\\
\bf{2^e_D\mapsto 4^m_D}
	&\qquad d*F^{\prime}_4= F^{\prime}_4\wedge H_3
	&\qquad d{\cal F}_2 =\hat H^{(5)}_3\\
& & \\ 
\bf{1^e_F\mapsto 6^m_D}
	&\qquad	d*H_3= F_2 \wedge *F^{\prime}_4
	&\qquad d\star {\cal F}_2 =*\hat F^{\prime (7)}_4\\
\bf{4^m_D\mapsto 6^m_D}
	&\qquad dF^{\prime}_4=  F_2 \wedge H_3
	&\qquad d{\cal F}_2 =\hat H^{(7)}_3\\
& & \\
\hline
& & \\
 \bf{0^e_D\mapsto 5^m_S}
	&\qquad d*F_2= H_3 \wedge  *F^{\prime}_4 
	&\qquad d\star {\cal G}_1 =*\hat F^{\prime (6)}_4\\
\bf{4^m_D\mapsto 5^m_S}
	&\qquad	dF^{\prime}_4= H_3 \wedge F_2
	&\qquad d{\cal G}_1 =\hat F^{(6)}_2\\
& & \\ 
\bf{2^e_D\mapsto 5^m_S}
	&\qquad d*F^{\prime}_4 =H_3 \wedge F^{\prime}_4
	&\qquad d\star {\cal G}_3=d{\cal G}_3=\hat F^{\prime (6)}_4\\ 
\end{array}
\]
\caption{Intersections   in type IIA}
\label{tableiia}
\myrule\vskip -.5cm
\end{table}

\begin{table}[h]
\myrule
\[
\begin{array}{ccc}
\bf{1^e_F\mapsto 3^*_D}
	&\qquad d*H_3=F_5 \wedge F^{\prime}_3
	&\qquad d\star {\cal F}_2 =\hat F^{\prime (4)}_3\\
\bf{1^e_D\mapsto 3^*_D}
	&\qquad d*F^{\prime}_3=F_5 \wedge H_3
	&\qquad d{\cal F}_2 =\hat H^{(4)}_3\\
& & \\ 
\bf{1^e_F\mapsto 5^m_D}
	&\qquad	d*H_3= F^{\prime}_3 \wedge F_5
	&\qquad d\star {\cal F}_2 =\hat F^{(6)}_5\\
\bf{3^*_D\mapsto 5^m_D}
	&\qquad dF_5= F^{\prime}_3 \wedge H_3
	&\qquad d{\cal F}_2 =\hat H^{(6)}_3\\
& & \\ 
\bf{1^e_F\mapsto 7^m_D}
	&\qquad	d*H_3= F_1 \wedge *F^{\prime}_3
	&\qquad d\star {\cal F}_2 =*\hat F^{\prime (8)}_3\\
\bf{5^m_D\mapsto 7^m_D}
	&\qquad dF^{\prime}_3= F_1 \wedge H_3
	&\qquad d{\cal F}_2= \hat H^{(8)}_3\\
& & \\
\hline
& & \\
\bf{1^e_D\mapsto 5^m_S}
	&\qquad  d*F^{\prime}_3= H_3 \wedge F_5
	&\qquad d\star {\cal G}_2 =\hat F^{(6)}_5\\
\bf{3^*_D\mapsto 5^m_S}
	&\qquad dF_5= H_3 \wedge F^{\prime}_3
	&\qquad d{\cal G}_2 =\hat F^{\prime (6)}_3\\
& & \\
\hline
& & \\
\bf{1^e_F\mapsto 1^e_D}
	&\qquad d*H_3= *F^{\prime}_3 \wedge F_1
	&\qquad d\star {\cal F}_2  =\hat F^{(2)}_1\\
& & \\
\bf{5^m_D\mapsto 5^m_S}
	&\qquad dF^{\prime}_3 = H_3 \wedge F_1
	&\qquad d\star {\cal G}_6 =\hat F^{(6)}_1\\ 
\end{array}
\]
\caption{Intersections  in type IIB}
\myrule\label{tableiib}
\end{table}

\begin{table}[h]\label{table11}\myrule
\[
\begin{array}{ccc}
\bf{2^e \mapsto 5^m}
	&\qquad d*F_4=F_4 \wedge F_4
	&\qquad d\star {\cal G}_3=d{\cal G}_3=\hat F^{(6)}_4
\end{array}
\]
\caption{Intersections in $D$=11 supergravity}
\myrule\end{table}

The intersections fall into
two classes. The first is associated with the intersection of the
fundamental string with D$p$-branes, its dual is always a D$(p-2)$-brane
\cite{str}.  The second class consists of intersections of D-branes with
the solitonic five brane $5_S$. These are the only branes which can end on
the $5_S$-brane.

Notice in Table \ref{tableiia} the ``opening'' of a D0-brane on the
D2-brane and on the solitonic 5-brane. It corresponds to an intersection of
dimension -1 which does not make obvious sense.  However one can easily see
that the relation \rref{intformula} stills holds for the D0-brane
world-line intersections in the Euclidean provided the electric fields are
given imaginary values.  Note also that in type IIA theory the
intersections with the D6-brane imply a 2-dimensional transverse space,
leading to a non-asymptotically flat behavior for the harmonic functions
determining the metric. A similar phenomenon occurs for the intersections
with the D7-brane in type IIB theory, where the transverse space is one
dimensional.  Nevertheless, one can show that formula \rref{intformula}
still holds in these two cases.

In type IIB theory, there exists a $SL(2,R)$ duality group acting on $\phi$
and $A_0$ by fractional linear transformations on the complex scalar $\tau
= A_0+ ie^{-\phi}$, i.e.
\[
\tau \rightarrow \frac{a\tau + b}{c\tau + d}\ .
\label{eq:fraclin}
\]
All the equations of motion and Bianchi identities are $SL(2,R)$ invariant
provided that the 2-form-valued vector $(A_2,B_2)$ transforms as an
$SL(2,R)$ doublet:
\[
\label{fraclinc}
\left(\begin{array}{c}A\\B\end{array}\right) 
\rightarrow
\left(\begin{array}{cc} a & b \\ c & d \end{array}\right) 
\left(\begin{array}{c}A\\B\end{array}\right).
\]

The last two cases of Table \ref{tableiib} are built up from objects
which are $SL(2,R)$-dual to each other. It follows that 
the configurations $\bf{1^e_D\mapsto 1^e_F}$ and $\bf{5^m_S\mapsto 5^m_D}$
exist by S-duality. In fact there is a continuum of dyonic configurations
\cite{schwarz}
obtained by acting with the $SL(2,R)$ group. These 
configurations do not have duals. Indeed, the dual configurations
would involve intersections with a (-1)-brane, which do not make   sense
even in the Euclidean.

In both type II theories, the opening of closed objects on D-branes is
permitted by the appearance of boundary charges coupled to the ubiquitous
$2$-form ${\cal F}_2$ which describes degrees of freedom of the brane. All
such Goldstone fields are related by dimensional reduction: reducing the
dimension of a D-brane by one results in the corresponding dimensional
reduction of ${\cal F}_2$, yielding the transmutation 
of one vector field component
to one scalar Goldstone field living in the dimensionally reduced brane and
describing its additional translational degree of freedom. On the other
hand, the openings on the solitonic five-branes require fields which can
be related only to D-brane world-volume duals of ${\cal F}_2$. In type IIA,
the solitonic 5-brane is characterized by two independent field strengths
living on its world-volume, a 1-form ${\cal G}_1$ and a self-dual 3-form
${\cal G}^+_3$. This field content, established by the open brane picture,
is consistent with the 11 dimensional origin of the 5-brane, the 0-form
potential being the 11th dimension.  Indeed, Table \ref{table11} shows that
there is a self-dual 3-form living on the world-volume of the 5-brane.  In
type IIB the fields are ${\cal G}_2$ and ${\cal G}_6$, the latter being a
non propagating field.  This world-volume field content of the solitonic
5-branes was derived in \cite{callan} by directly identifying the Goldstone
modes of the supergravity solution.
 
\section{Conclusion}

We have shown that all the configurations with a closed
$p_a$-brane intersecting a closed $p_b$-brane on a $p_a-1$ dimensional
intersection with vanishing binding energy can be opened.  Charge
conservation and gauge invariance are simultaneously ensured by Goldstone field
strengths appearing on the world-volume of the $p_b$-brane.  Each such
intersection has a partner such that their respective boundaries carry
dual charges. Note that the field strengths in the branes
have been generated from supergravity without any
appeal to the apparatus of string theories.

A particular case is that where a fundamental string opens on a
D-brane. One can then consider the reverse process where open strings
collapse to form closed strings with zero binding energy to the D-brane.
Energy conservation permits the closed strings to separate from the
brane. This process can be viewed as the classical limit of the well known
quantum process of closed string emission by D-branes \cite{emission}.
Our analysis shows, at the classical level, that this string process can be
extended to the emission of closed branes. Any closed brane supports
traveling open branes attached to it.  These configurations of open branes
break supersymmetry and can collapse through annihilation of boundary
charges into configurations containing closed branes with zero binding
energy. This leads to the emission of supersymmetric closed branes.

The generality of this phenomenon points towards the existence of an
underlying quantum theory where the emission of closed branes and that of
closed strings occur as similar phenomena.

We have seen that all D-branes carry fields ${\cal F}_2$ related
to each other through dimensional reduction. The reason behind this
universality is that all D-branes can receive charges from string
boundaries.  But this is not the case for the solitonic five-branes, who
can host only dual charges stemming from D-brane boundaries.  In this
sense, string emission and D-brane emission are dual processes.

The supersymmetric $U(N)$ matrix theory attempt to describe a universal
``M-theory'' of strings and branes can be viewed, in the classical abelian
limit, as a common dimensional reduction of all Goldstone ${\cal F}_2$
fields to the extreme cases of $0$-branes \cite{matrix1} or $-1$-branes
\cite{matrix2}. It may appear therefore as a candidate to treat
democratically strings and branes.  However the solitonic five-branes
excitations as well as D-brane emission from D-branes or from solitonic
branes, are related to ${\cal F}_2$ by dualities in various
dimensions. These features can hardly fit into this scheme in a
straightforward way.

Whatever the outcome of the matrix approach will ultimately be, the
discretization of space (or space-time) related to its matrix description
is welcome. Indeed, classical continuous $p$-branes for arbitrary $p$ do
not appear to admit a consistent quantization and should arise as an
approximate but not as an exact solution of the theory.

\subsection*{Acknowledgments}

L. H. would like to thank the members of the {\it Centro de Estudios
Cientificos de Santiago} and P.W. the members of the {\it Service de
Physique Th\'eorique} of ULB for their warm hospitality during the
completion of this work. This work was supported in part by the
{\it Centre National de la Recherche Scientifique} and the EC TMR
grant {\sc erdbchrxct 920035}.

\end{document}